\documentclass[a4paper,9pt]{extarticle}
\usepackage[utf8]{inputenc}
\usepackage{amssymb}
\usepackage{appendix}
\usepackage{lscape}
\usepackage{graphicx}
\usepackage{amsmath}

\usepackage{biblatex}
\usepackage{setspace}

\usepackage[utf8]{inputenc}
\usepackage{geometry}
\title{A Deep Learning Method for Predicting Mergers and Acquisitions: Temporal Dynamic Industry Networks}
\author{Dayu Yang\\{\small University of Delaware}}
\date{}

\begin{document}

\maketitle

\begin{abstract}
   \noindent Merger and Acquisition (M\&A) activities play an increasingly crucial role in market consolidation and restructuring. For companies pursuing acquisitions, M\&A represents a significant investment strategy, with one of the primary incentives being the attainment of complementarities to enhance market power within competitive industries. Beyond the intrinsic factors extensively studied in existing empirical research, a firm's M\&A behavior is actively influenced by the M\&A activities of its peer firms---a phenomenon known as the ``peer effect.'' However, current research often fails to capture the rich interdependencies of M\&A events within the industry network.

An effective M\&A predictive model should provide fine-grained, deal-level predictions without requiring ad-hoc feature engineering or data rebalancing. Such a model can deliver precise predictions of rival firms' M\&A behaviors and offer specific M\&A recommendations for bidder and target companies. Most existing models, however, predict the behavior of only one side of M\&A deals, lacking the capacity to make firm-specific deal recommendations. Additionally, they often rely on ad-hoc data truncations that convert continuous time information into arbitrary intervals, leading to information loss and impairing predictive performance. The laborious and arbitrary nature of these transformations also hampers the applicability of these models in real-world scenarios. Moreover, due to the sparsity of M\&A events, existing deal-level predictive models often require ad-hoc data rebalancing, which introduces bias and further compromises performance and reliability.

To address these limitations, we propose a novel M\&A predictive model based on Temporal Dynamic Industry Networks (TDIN), leveraging temporal point processes and deep learning techniques. Our approach captures the rich interdependencies among M\&A events without the need for ad-hoc feature engineering or data rebalancing, as the temporal point processes inherently model event sparsity within their intensity functions. Empirical evaluations on M\&A cases from January 1997 to December 2020 demonstrate the superiority of our method in predicting M\&A activities and providing actionable deal-level recommendations.\footnote{Code is available at: \url{https://github.com/dayuyang1999/Merger_Acquisition_Prediction}}

\end{abstract}

\section{Introduction}

Merger and acquisition (M\&A) is a transaction in which the ownership of a company is transferred to another company. A M\&A transaction contains three important elements: the time of the transaction, the acquirer, and the target, where the target come from a target candidate set determined by the acquirer (Gugler \& Konrad, 2002).

\begin{figure}
    \centering
    \includegraphics[scale=0.3]{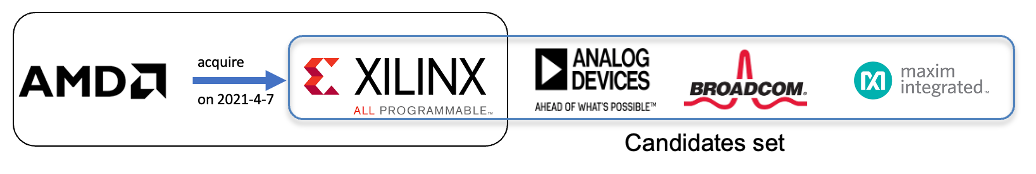}
    \caption{An example of merger and acquisition events}
    \label{fig:my_label}
\end{figure}

M\&A is a driving force for market consolidation and reconstruction. It also plays an increasingly important role in the US and global economy. According to Thomas Reuters’ Global Mergers \& Acquisitions Report, M\&A events value about 6 percent to 10 percent of the entire US GDP in the recent two decades. The value of worldwide M\&A activities in 2021 has reached 3.6 trillion US Dollar until August. This number has surpassed the total number of 2020.  

There is a broad list of theories explaining the motives behind merger and acquisition events (Trautwein, 1990), but one factor has long been viewed as a key driver of mergers and acquisitions: pursuing complementarity. In a successful deal, the acquirer first identifies suitable target candidates which can provide potential complementarity, and then selects the most favorable one from the candidate set (Gugler \& Konrad, 2002). One important benefit of obtaining complementarity is: the capability of creating new or better products compared with the peer firms, which in turn increases the market power of the acquirer (Hoberg \& Philips, 2010). The target selection process is a critical step for every M\&A event, because an inappropriate choice will very likely lead to the failure of a bid eventually. 

Many empirical studies have studied that intrinsic factors of firms that could significantly affect the probability of forming a successful M\&A deal. Those intrinsic factors are financial or/and managerial variables of both/either acquirers and/or targets suggested by various financial hypotheses. For example, firm size, cash flow, market-to-book ratio, etc, have been well-documented that they are effective predictors of successful M\&A deals.

Besides financial and managerial variables, recent studies (Hoberg \& Phillips, 2010; Ahern \& Harford, 2014; Bernard et al., 2020; Bustamante \& Fresard, 2021) start to explore interdependencies among M\&A events in the local environments around focal firms. Following the motivation of M\&A: obtaining complementarity, the probability of completing a successful deal relates to whether the target has assets, skills, or technologies that can help the acquirer improve or differentiate their products from their rivals (Hoberg \& Philips, 2010). In the meanwhile, firms with close peer firms must compete for restructuring opportunities. That is, historical M\&A events by its peer firms can affect a focal firm's M\&A intentions due to the increase of competitive level of the local environments. Hoberg \& Philips (2010) use an indicator to represent such competition among peer firms. Bustamante et al. (2021) found theoretical and empirical evidence that firms actively track the investment behavior of their peer firms and react to these behaviors. Ahern \& Harford (2014) found that such “peer influence” can propagate contagiously through industry networks and make further impact on those firms that are not located in the local environments, which form the well-documented pattern of M\&A events: merger waves. Further, the local environment of a firm is not stationary, which brings more complexity to the analysis of M\&A deals. Firms are continuously evolving as their intrinsic factors vary over time, the local environments are also changing frequently (Hoberg \& Philips, 2016).

Based on existing empirical literature, we can roughly classify the factors that could possibly influence the incentive of an acquirer to trigger a new merger and acquisition in the future to two categories: intrinsic and extrinsic factors. The intrinsic factors are the inner initiative from the acquirers themselves. They mainly include the financial and managerial factors indicated by previous literature. While extrinsic factors describe the cumulative “peer effect” (Bernard et al., 2020; Bustamante \& Fresard, 2021) from the past M\&A events triggered by the peer firms.

M\&A prediction is an important task and has received a lot of research attention (Palepu, 1986; Barnes, 1999; Powell, 2004; Brar et al., 2009; Xiang et al., 2012; Yang et al., 2014; Yan et al., 2016; Routledge et al., 2017; Moriarty et al., 2019). For business researchers, a M\&A predictive task helps them discover the underlying driving factors of M\&A events, and provides a method to verify financial hypotheses. A M\&A predictive task is important for firm managers as well. It helps firms to predict the future investment behavior of their rivals and therefore allows them to make more prompt decisions (Bernard et al., 2020; Bustamante \& Fresard, 2021).  It is also important to firms seeking targets to merge with. A M\&A predictive model can learn the implicit pattern of historical M\&A events and recommend appropriate target candidates who provide potential complementarity (Trautwein, 1990; Shleifer \& Vishny, 2003). On the other hand, a M\&A predictive is also important for firms seeking buyers such as startups. M\&A prediction might help those firms to identify potential buyers and assess their probability of being successfully acquired. For investors including venture capitalists and portfolio managers, by making a prediction of which firm will get involved into a complete M\&A deal in the future, a M\&A predictive model helps them identify firms has high growth potential and harvest acquisition premium (Brar et al., 2009).

In terms of the classification label, there are two categories of predictive works: single-side and deal-level predictions. Most existing studies lie in the first category. Their predictive labels only focus on a single side of a complete M\&A transaction: “if a firm will be involved in a M\&A transaction in the future as a target or acquirer”. And they usually solely consider the information about one side of the transaction as predictors. Such works fail to predict precise investment behavior due to the lack of prediction of one side. Secondly, they cannot help firms to identify their corresponding potential targets or acquirers. Similarly, for investors, the predictions made by these M\&A predictive works is unable to provide potential target firms that are specific for them. Also, a M\&A transaction is conceptually a mutual agreement between acquirer and the target. Those works fail to take advantage of the information of both sides of a transaction which may harm the predictive performance. Another category of predictive work is called a deal-level M\&A predictive task. Deal-level predictive tasks consider both the information of acquirers and targets as their predictors. Their predictive label are usually: “if an acquirer-target pair could complete a M\&A transaction in the future”. They are generally more attractive due to the overcoming of the former-listed drawbacks. However, deal-level prediction tasks face a big challenge: the training data is highly imbalanced. For example, creating a random acquirer-target pair, it is very unlikely that this firm pair could finally form a successful deal due to the sparsity nature of M\&A transactions (Yang et al., 2014). To be clear, a deal-level prediction task can be fundamentally viewed as a link prediction task in a dynamic industry network with two objectives. The first is the time that a new directed link is created between two firms. The second one is which two nodes are involved in the new directed link.

Focusing on the model structure, most existing works use acquisition likelihood models (Palepu, 1986) to make predictions. It is a logistic regression that considers financial, managerial, technological or other variables as predictors. The issues of logistic regression are: first, the linear assumption was made between predictors and the log odds. Also, only a summation form can apply between independent variables, which only enables the restrictive representation of the relationship between related features of a M\&A transaction. Moreover, regression methods dismiss the precise timestamp such that the original time information is transformed to yearly features. For example, A M\&A event happened in January 1st, 2020 and another one happened in December 31st, 2020 will be both marked as "2020". Although the latter M\&A event is possible to be closer with a M\&A event happened in 2021. At the same time, this type of predictive is unable to make fine-grained M\&A predictions that allows the investors and managers to make precise business decisions. Also, existing M\&A predictive work fails to model the rich inter-dependencies of M\&A events. They either did not consider them, or only represented the inter-dependencies via highly restrictive formations. Specifically, for most predictive work that did not make deal-level prediction, they failed to consider the conceptual foundation of a merger and acquisition event as a mutual agreement between two sides, and therefore did not consider the rich information from one side of the deal. This brings error to the estimation of coefficients in empirical analysis due to the ignorance of strongly related variables: the error terms may not be white noise. For the existing deal-level prediction works, they handle the sparsity issue by implementing resampling techniques. This may bring bias to the model. And the model may finally yield biased outcomes since certain patterns may only exist under the condition that data has been oversampled. If we borrow the method from dynamic link prediction works, the rich inter-dependencies among historical M\&A events may be captured. However, existing link prediction models are hard to apply to M\&A since the M\&A event induced network is large-scale, dynamic and sparse.

To address the challenges above-mentioned, we propose a novel Temporal Dynamic Industry Network (TDIN) that describes each firm as a node in the network, while the structure of the network is determined by their prompt competitive relationships instead of the interaction between nodes. Our model is able to capture the rich inter-dependencies among historical merger and acquisition events, and make continuous-time deal-level predictions. Our model also solves the sparsity issue by modeling the sparsity into the parameter. Compared with existing dynamic network methods, our model does not require dense interactions among the nodes to make precise predictions.

\vspace{20pt}

\section{Literature Review}

Our work is related to two streams of research: existing M\&A predictive works and deep learning based dynamic link prediction methods. In the following subsections, we will first give a review of each stream and discuss the main novelties of our study.

\subsection{M\&A Predictive Works}

Most of the existing M\&A predictive works employed acquisition likelihood model (Palepu, 1986) to predict the probability of a M\&A predictive label. The choice of predictive label varys.

$$P(Y=1) = \frac{1}{1+e^{-\beta x}}$$

Those predictive works set $P(Y=1)$ to be the probability that “a firm will be involved in a M\&A event as target in the next year” (Palepu, 1986; Barnes, 1999; Powell, 2004; Brar et al., 2009; Xiang et al., 2012; Routledge et al., 2017; Moriarty et al., 2019). At the same time, a small portion of the predictive work set P(Y=1) to be “the probability that a firm will be involved in a M\&A event as acquirer in the next year” (Yan et al., 2016; Routledge et al., 2017; Moriarty et al., 2019). These two groups of works are called "single-side" predictive works. There are pros and cons for both of the options. For those works which set target as the predictive label in the acquisition likelihood model, there will be more training samples for the likelihood model, and therefore obtain the estimation of coefficients with lower variance (Brain et al 1999). This is because a single acquirer usually triggered multiple M\&As while, generally speaking, a target was usually acquired by only once (Although it is possible that the ownership of a target was transferred among different acquirers many times). However, the disadvantage of this predictive label setting is that the accessible information of acquirers are usually richer, since the acquirers are usually larger companies (Meador et al., 1996; Barnes, 2000; Tsagkanos et al., 2007; Routledge et al., 2017) that have been existing in the market for a long time (Tsagkanos et al., 2007), while, in general, targets are relatively smaller (Palepu, 1986) and start-up firms (Tsagkanos et al., 2007). 

Although there are a dominant number of the predictive works which set the predictive label as “if a firm will be involved in a M\&A event as target/acquirer in the next year?”. The most attractive and challenging predictive task is the deal-level prediction. Conceptually, a complete M\&A event is indeed a mutual agreement between the acquirer and the target (Blake et al., 1985), while the first two categories of prediction models only considered one side of a complete M\&A transaction. To the best of our knowledge, the only deal-level predictive task introduced so far is a M\&A prediction task specifically targeting the technological industry (Yang et al., 2014). 

However, deal-level prediction faces a big challenge: the extremely imbalanced data. Since the M\&A is an infrequent investment event for companies. If comprehensive matchings of firm pairs are considered. The number of positive acquirer-target samples will be much less than negative samples. This leads the model to focus more on classification of major samples while ignoring or misclassifying minority samples. However, in this case, the minority samples are merger and acquisition events we are trying to predict. 

The acquisition likelihood model, or logistic regression, dominates the M\&A predictive task. In sum, the major difference between those studies are the variables they used as regressors. The most common variables are financial and managerial variables indicated by theories of merger motives (Trautwein, 1990). They usually include firm size, market-to-book-value ratio, cash flow, return on assets, sales to total assets (asset turnover), debt-to- equity ratio, price-to-earnings ratio, current ratio, return on equity, debt-to-assets ratio, capital-expenditures-to-total-asset ratio, and growth (Yang et al., 2014). While some other studies show the predictive power of other innovative variables, such as: technological quantity (Yang et al., 2014) and frequency of words and phrases (Routledge et al., 2017).

Recent studies (Hoberg \& Phillips, 2010; Ahern \& Harford, 2014; Bernard et al., 2020; Bustamante \& Fresard, 2021) start to put attention into how local environments of firms affect the likelihood of future M\&A events. They propose that the local environment affects M\&A deals in various ways. For example, since mergers and acquisitions are a quick way to obtain complementarity and then allow the companies to create new or better products that differentiate themselves from their peer firms, the product of those peer firms can affect the choice of target. The target that is finally acquired by the acquirer should provide a great amount of complementarity and therefore enable the acquirer to largely differentiate itself from its peer firms after the M\&A deal is completed (Hoberg \& Phillips, 2010). Hoberg and others construct the variable “the fraction of its ten closest peer firms that were either targets or acquirers in the previous year” to model the influence of historical M\&A events triggered by peer firms. In the meanwhile, firms are constantly grasping the M\&A events triggered by their peer firms, and actively respond to the investments (Bernard et al., 2020; Bustamante \& Fresard, 2021). This means that the historical M\&A events triggered by focal acquirer’s peer firms could still affect the focal acquirer’s incentives to trigger a new M\&A event. Moreover, such “peer influence” can propagate contagiously through industry networks to make further impact on those firms that are not located in the local environment, and form a well-documented pattern of M\&A events: merger waves (Ahern \& Harford, 2014). Specifically, Ahern et al. used several lag terms to describe the influence of historical events in the local industry network.

Besides the acquisition likelihood models, Xiang et al. used Latent Dirichlet Allocation to predict M\&A events for startups. They further introduced Bayesian networks for the prediction. Moriarty et al. also used Latent Dirichlet Allocation method to extract several topics from 10-K filings\footnote{A 10-K is a comprehensive report filed annually by a publicly-traded company about its financial performance and is required by the U.S. Securities and Exchange Commission (SEC).}, then used the topic embedding vectors to make M\&A predictions. Yan et al. proposed a multi-dimensional Hawkes process to make M\&A predictions. Their model design was motivated by the well-documented “merger waves” phenomenon. A Hawkes process (Hawkes, 1971) is a self-motivating process that could mimic such a pattern in M\&A history.

\subsection{Temporal Point Processes}

Temporal point process (TPPs) is a basic framework that models discrete event sequences in a continuous-time space. Compared with other sequence models such as time-series models and recurrent neural networks (RNNs), temporal point process allows us to keep the original precise information of time: RNNs and time-series can only model sequence in a discrete space while transform the original timestamps in continuous space to "stages" in discrete space. For some dynamic predictive tasks such that we only care about how the object will evolve in the next stage, RNNs and time-series models are suitable since the we do not need a precise prediction for time. However, for mergers and acquisitions, investors and companies benefits a lot from the precise prediction about the investment target firms or their rival firms behaviors. Under this problem, temporal point processes are a better framework to start with, let alone the remarkably well-established theoretical foundation TPPs have compared with recent deep learning sequence models such as recurrent neural network (Rumelhart, 1986) and transformer (Vaswani et al., 2017). The keystone of a temporal point process is its (conditional) intensity function i.e. the average number of events that happen in a unit time. Formally, in an infinitesimal time interval $dt$, let $N(t)$ be the total number of events happened until $t$. So the number of new event happened in the time interval $dt$ will be $N(t+d t)-N(t)$. Let $\lambda(t)$\footnote{Sometimes, people represent conditional intensity as $\lambda^*(t)$, where the star * represents "conditional" (Daley \& David, 2007).} be the value of intensity function conditioned on historical M\&A events observed until $t$, $\mathfrak{H}_t$ (up to but not including time $t$), we have the following definition:

\vspace{10pt}

$$\lambda(t)=\lim _{d t \rightarrow 0} \frac{\mathbb{E}\left(N(t+d t)-N(t) \mid \mathfrak{H}_{t}\right)}{d t}$$

\vspace{10pt}

where $\mathbb{E}\left(N(t+d t)-N(t) \mid \mathfrak{H}_{t}\right)$ is the expected number of new events happened in the infinitesimal interval $d t$, conditioned on historical observations $\mathfrak{H}_t$. 

Under the assumption of regular point processes that most recent studies imply, the probability that multiple(more than one) event happened in a same infinitesimal time interval is a higher-order infinitesimal value with the length time interval (roughly speaking, the probability is 0). Then we have another equivalent definition of intensity function:

$$\lambda(t)=\lim _{d t \rightarrow 0} \frac{\mathbb{E}\left(N(t+d t)-N(t) \mid \mathfrak{H}_{t}\right)}{d t} = \lim _{d t \rightarrow 0} \frac{\operatorname{Pr}(N(t+d t)-N(t)=1)}{d t}$$
This means the probability that there is a new event happening in this infinitesimal time interval $dt$ equal to the Expectation of the number of new events happening in the same infinitesimal time interval $dt$.

The intensity function (or conditional intensity function) is the key to every point processes since it controls the behavior of the event sequences. In practice, we should formulte a specific parameterization form of intensity functions that allows the point processes to fit our observations. Recent studies have generalized point process to neural point processes (Mei \& Eisner., 2017; Du et al., 2016). The difference between neural point processes and traditional point processes such as Poisson processes, Hawkes processes is that the intensity function of a neural point process is parameterized by a deep learning method, which abandoned the explicit parametric formulations of intensity functions. Although the model capacity (often positively related to prediction accuracy) has been significantly increased after introducing deep learning methods into point processes, some theoretical conclusions may be no longer exists (Mei \& Eisner., 2017). For example, the likelihood function may no longer have an analytical solution, which means we need to use numerical methods to estimate some values (Du et al., 2016). This may harm the predictive performance of point processes.

Most point process uses maximum likelihood estimation (MLE) to estimate its parameters. For a regular point process (Robin, 1972), assume we observed a event sequence happened at $t_1, t_2, ..., t_n$. The joint density function can be written by:

$$
    f\left(t_{1}, t_{2}, \ldots, t_{n}\right)=\prod_{j} f\left(t_{j} \mid \mathfrak{H}_{j}\right)$$

where $\mathfrak{H}_{j}=\left(t_{1}, t_{2}, \ldots, t_{j-1}\right)$ is the historical events happpened before the timestamp $t_j$ (but not including this timestamp).

We have the relationship between the intensity function and probability density function\footnote{proof see Appendix}:

\begin{equation}
    f\left(t_{j+1}\right)=\lambda\left(t_{j+1}\right) \exp \left(-\int_{t_{j}}^{t_{j+1}} \lambda(\tau) d \tau\right)
\end{equation}

Accordingly, the log-likelihood function can be written by:

$$
    \log f\left(t_{1}, t_{2}, \ldots, t_{n}\right)=\sum_{j=1}^{n} \log \lambda\left(t_{j}\right)-\int_{t_{0}}^{t_{n}} \lambda(\tau) d \tau
$$

Many events in the world are correlated to each other. One type of temporal point processes that can model a specific "correlation" is the Hawkes process (Hawkes, 1971). For a vanilla Hawkes process, its conditional intensity is written by (Hawkes, 1971):

\begin{equation}
 \lambda(t)=\mu+a \sum_{i: t_{i}<t} g\left(t-t_{i}\right)   
\end{equation}

where the current timestamp is $t$. $t_i$ is the timestamp that most recent event happened before $t$. $\mu$ is called "the spontaneous rate" or "the base rate". $g(t)$ is a monotonous decaying function with time, which models the phenomenon that "the influence from the past events is decaying over time". In the original paper, Hawkes use an exponential decay function due to several advantages (Hawkes, 1971). From equation (2), we could find a Hawkes process assume: First, all historical events can influence the later events. Second, the influence could only be additive and positive.

Hawkes process also exists in multi-dimensional form. The multi-dimensional hawkes process was called mutually-exciting Hawkes process since it assumes the event triggered by any dimension can effect any other dimensions. Yan et al. (2016) models M\&A by a multi-dimensional Hawkes process. The reason for that is Hawkes processes can simulate the wave-like phenomenon for M\&A events. However, according to the previous empirical studies (Hoberg \& Phillips, 2010; Ahern \& Harford, 2014; Bustamante \& Fresard, 2021), only the M\&A events triggered by peer firms can affect the future M\&A decisions of the focal firms, while Yan et al. (2016) assume that the M\&A event triggered by any firm can affect any other firms regardless of the similarity or distance between firms. Yan et al. (2016) also assume such effect must be positive.

% add multi-dimensional and marked?

As we shown above, regular temporal point processes such as Hawkes processes usually draw strong parametric assumptions (Du et al., 2016; Mei \& Eisner, 2017) about the intensity function. This leads to an issue that those intensity functions may not be able to fit the events that we observe very well. Mathematically, this leads the value of the cumulative likelihood to be very low regardless of the various training techniques. For events sequence like mergers and acquisitions, events are related to one another via a very complex mechanism. The strong parametric assumptions of traditional point processes may not be capable of fitting the complicated behavior of event sequence. Thanks to the recent development of deep learning methods, recent studies have started to combine the deep learning methods with temporal point processes (Du et al., 2016; Mei \& Jason, 2017; Li et al., 2018; Omi et al., 2019)

Similar to the traditional TPP models, maximum likelihood has been adopted as the learning objective in many network based models (Du et al., 2016; Mei \& Eisner, 2017). Given the sequence set ${S^i}$ for $S^i = (t_j^i, y_j^i)_{j=1}^{n_i}$, by assuming the independence of the event type and timestamp, the objective can be simplified by the factorized model:

For traditional and neural temporal point processes, most previous works (Du et al., 2016; Mei \& Eisner, 2017) learn the parameter via maximum likelihood estimation (MLE):

\begin{equation*}
    max \;L = \prod_{i=1}^{n} f(t_n) = [\prod_{i=1}^{n} \lambda (t_i)] \exp(-\int_0^{t}\lambda(\tau) d\tau)
\end{equation*}

There are some works model multiple types of events into a single sequence, which is called "marked point processes" (Carter \& Prenter, 1972; Ripley \& Kelly 1977; Ripley 1997). If we assume the independence of the event type(or mark) and time, the objective of MLE can be separated into two independent terms, which compute the loss of mark and time separately (Du et al., 2016):

\begin{equation*}
    max\;\sum_{i} \sum_{j}\left(\log P\left(y_{j+1}^{i} \mid \boldsymbol{h}_{j}\right)+\log f\left(d_{j+1}^{i} \mid \boldsymbol{h}_{j}\right)\right)
\end{equation*}
    
where $log P(y_{j+1}^{i}|h_j)$ is the time loss computed by likelihood loss. Since predicting type of events can be viewed as a multi-class classification problem, $log f(d_{j+1}^{i}|h_j)$ is the type loss that is computed via cross entropy loss function.Two major difference between (Du et al., 2016) and (Mei and Eisner, 2017) is that: first, the varying of time is modeled by a single parameter in (Du et al., 2016), this allows Du et al., to derive the analytical solution for the density function. On the contrary, in (Mei and Eisner, 2017), the authors model the varying of time implicitly in the model strucutre, which leads to the problem that the maximum likelihood function has no analytical solution. Mei et al. had to use Monte-Carlo estimation to estimate the likelihood, which is computationally expensive compared with (Du et al., 2016). Second, the former uses one intensity function for all types while the latter allocates respective intensity functions to each event type.

\subsection{Graph Neural Network}

A deal-level M\&A predictive problem can be fundamentally viewed as a directed link prediction problem in a dynamic industry network, where there are two objectives: timing that a new link formed, and which two nodes are involved in the new link. For example, Yan et al. (2016) modeled merger and acquisition events in a fully-connected network following the naive assumption that: a historical M\&A event triggered by a firm can influence the future M\&A decisions of all other firms existing in the network. Hou et al. (2015) also model M\&A events from a complex network perspective, and state that the cascading-failure phenomenon is caused by the contagious behavior of nodes in the beer industry network which they called “coupling mechanisms”.
Recently, graph structured data is well studied since its broad applicability. Except M\&A events induced network, social network, epidemiological network, physical network, among others are all graph unstructured data that is hard to be modeled by traditional models. A graph $\mathcal{G}= (\mathcal(V), \mathcal(E))$ is comprised of a set of $N$ nodes, $\mathcal(V)$, and the edge, $\mathcal(E)$, between them. In our case, a M\&A events induced network may consist of a set of firms (nodes), and the edges may indicate whether two firms have formed a transaction deal with each other. Although unstructured data is hard to be modeled by traditional models such as linear model, tree model, etc, graph neural networks (GNNs) (Scarselli et al., 2009; Li et al., 2015; Bruna et al., 2013; Kipf \& Welling, 2016; Santoro et al., 2017; Hamilton et al., 2017) have shown promising performance in link prediction problems and representation learning problems. These models use recursive neighborhood aggregation to learn latent features, $Z^{(t)}$ of nodes given the feature of the last state, $Z^{(t-1)}$.

$$Z^{(t)}=f\left(Z^{(t-1)}, A, W^{(t)}\right)$$

Where $A \in \mathbb{R}^{N \times N}$ is an adjacency matrix of graph $\mathcal{G}$. $W^{(t)}$ are trainable parameters; $d, c$ are input and output dimensions, and $f$ is a permutation invariant function, which is typically an aggregate function followed by a nonlinearity. State $t$ can correspond to either a layer in a feedforward network or a time step in a recurrent neural network (Li et al., 2015).

In sum, graph neural network (GNN) we used here is a family of graph learning algorithms that model the graph structure in the euclidean space (Hamilton et al., 2017b). Every GNN has two key components: Aggregate and Combine. For the focal node, a permutation invariant function take its neighbors' embeddings as input in the aggregation step. And then aggregated them as a node feature in the current recursive step. This node feature is further pass into a learnable tranformation layer(fully connected layer and recurrent neural network are the most widely-used ones). And we take the output of this layer as the new node embedding of the focal node.

The receptive field can be increased by stacking aggregation and combination mechanisms. If we have 2 layers of them, for every node, it can aggregate the information over 2-hop neighbors. However, generally speaking, the number of layers cannot be larger than 2 \footnote{see section 4.4 of (Hamilton et al., 2017) for detail}. Aggregate and combine components are also called "message passing mechanism". Formally, consider the graph $G=(V, E)$, where $V$ is the set of nodes. The feature matrix of nodes is $X$. The aggregate and combine components at layer $l$ can be represented by:

$$a_v^{l} = Aggregate^{l} (\{ h_u^{l-1}\; \forall u \in \mathcal{N}(v) \})$$

$$h_v^{l} = Combine^{(l)}(h_v^{l-1}, a_v^{(l)})$$

where $a_v^{l}$ is the node feature of $v$ that aggregated from it neighbors $\mathcal{N}(v)$ at layer $l$. Aggregate function can be chose from the set of permutation invariant functions, such as mean, max, min, etc. $h_v^{l}$ is the node embedding at layer $l$. It is the ultimate output of the message passing mechanism after the computation of layer $l$. Combine function is usually a fully connected layer that is composed by a linear transformation and a non-linearity.

Graph convolutional networks (GCNs) (Kipf \& Welling, 2016) is a GNN method that makes use of the symmetrically normalized graph Laplacian to compute the node embeddings. For a single GCN layer, the node features fro mthe previous layer is taken as input of the next layer, which can be seen as a Laplacian smoothing (Li et al., 2018). The recursive formula of GCN is introduced as:

$$H = f(\tilde{D}^{-\frac{1}{2}} \tilde{A} \tilde{D}^{-\frac{1}{2}})$$

Where $\tilde{A} = A + I_n$ and $\tilde{D}_{ii} = \sum_j \tilde{A_ij}$. $W$ is the parameter matrix. However, sometimes, having the structure information of the entire graph is impossible in advance. In a GCN, this leads to the issue that we are unable to compute the symmetrically normalized graph Laplacian. One solution is to interpret GCNs from the perspective of a single node. Borrowing the idea from (Xu et al., 2018a), the aggregate step of a GCN can be seen as:

$$a_v^{l} = Mean(\{ h_u^{l-1}\; \forall u \in \mathcal{N}(v)\cup u \})$$

This means, for a single node $v$, the algorithm is taking the average of its neighbors' node embeddings along with its own node embedding (a self-loop) to compute the aggregated node feature. This interpretation makes GCN an inductive graph neural network.

GraphSAGE (Hamilton et al., 2017) is a general inductive framework that learns node embeddings by sampling and aggregating neighbors' information. GraphSAGE, unlike GCN, is an inductive learning algorithm from the start which does not require the entire graph structure during training. The key innovation of GraphSAGE is sampling, which overcome the issue that: for some nodes, so many other nodes are connected with them. Sampling before aggregation largely increase the effciency of learning for GNN algorithms in extremely large graphs. The general graph convolutions implemented by GraphSAGE is:

$$\mathbf{h}_{v}^{(k)}=\sigma\left(\mathbf{W}^{(k)} \cdot f_{k}\left(\mathbf{h}_{v}^{(k-1)},\left\{\mathbf{h}_{u}^{(k-1)} \forall u \in S_{\mathcal{N}(0)}\right\}\right)\right)$$

where $h_v^{(0)} = x_v$, $f$ is an aggregate function. The aggregation function should be invariant to the permutations of node orderings, such as a mean, sum, or max function. $S_{N_{(v)}}$ is a random sample of the node v’s neighbors. $W$ is a learnable parameter, while the information aggregated from neighbors will be multiplied with $W$ to make a linear transformation. $\sigma$ is a nonlinear function which is generally seen as an activation function. $h_u^{(k-1)}$ is the node representation of node $u$ in $k-1_th$ recursion. Generally, the number of recursions should not surpass two. The node representations learned could further be used to make link predictions.

Graph Attention Network (GAT) (Velickovic et al., 2017) uses an attention module to compute the weighted mean of the neighbors' node embeddings. 

$$\mathbf{h}_{v}^{(k)}=\sigma\left(\sum_{u \in \mathcal{N}(v) \cup v} \alpha_{v u}^{(k)} \mathbf{W}^{(k)} \mathbf{h}_{u}^{(k-1)}\right)$$

$$\mathbf{h}_{v}^{(k)}=\sigma\left(\sum_{u \in \mathcal{N}(v) \cup v} \alpha_{v u}^{(k)} \mathbf{W}^{(k)} \mathbf{h}_{u}^{(k-1)}\right)$$

where the attention weight $\alpha_{vu}^{(k)}$ measures the connective strength between the node $v$ and its neighbor $u$. $g$ is a LeakyReLU activation function in the original paper (Velickovic et al., 2017). $a$ is a vector of learnable parameters. GAT further performs the multihead attention to increase the model’s expressive capability, especially on link prediction problems.

The static graphs, graphs with fixed adjacency matrix $A$, have been well studied and previous works in GNNs have proposed a significant number of methods that can apply in various scenarios (Bronstein et al., 2017). Among them, GraphSage (Hamilton et al., 2017) is an inductive method for learning functions to compute node representations that can be generalized to unseen nodes. Most other approaches only work with fixed graphs, which means if a node is introduced in the future, the node representations of all nodes in the network have to be re-trained completely. Just like the relationships between firms continuously change over time (Schmidt, 1975), in many cases, the network structure, or the corresponding adjacency matrix which embedded all the network structure information are not stationary (Bansal et al., 2010; Zhu et al., 2016; Goyal et al., 2017; Zhou et al., 2018; Minderer et al., 2019; Sun et al., 2019). The link prediction in an evolving network is called dynamic link prediction. There are several recent works that have explored the dynamic link prediction problem. In DynGem (Goyal et al., 2018), they borrowed the idea from autoencoder and constructed a graph autoencoder which can produce low-dimensional representations to represent the structure information over the entire network in a particular timestamp. Temporal attention is introduced by DySAT (Sankar et al., 2018), where the model learns the evolving graphs as a sequence of graphs. And the difference between graphs in the sequence is evaluated by a reconstruction error term. Other common sequence models, recurrent neural networks (RNNs) in deep learning are also introduced to model the temporal dynamics of a graph (Chen et al., 2018; Taheri et al., 2019). DyRep (Trivedi et al., 2019) is a method that learns the continuous-time dynamics of nodes based on temporal point processes. It supports two kinds of graph evolutions, structural change and communication between nodes. There are some dynamic link prediction works that assume the graph  $\mathcal{G}$ is unknown or unobservable (Yuan et al., 2017; Xu et al., 2019; Kipf et al., 2018). Among them, NRI proposes a general framework for learning representation from dynamic implicit graphs (Kipf et al., 2018).

\subsection{Key Novelties of Our Study}

The conclusions and results of the previous M\&A studies are beneficial. Palepu (1986) introduces the acquisition likelihood model based on logistic regression that is able to make predictions as "if a firm will be involved in a M\&A event in the future". Yang et al. (2014) gives the first deal-level predictive study that takes the information of both the acquirer and the target into account. They also provide a feasible solution for the sparsity issue of the M\&A events: resampling. Yan et al. (2016) uses a powerful statistical tool: point process to model M\&A events, and tries to reproduce the merger wave that is well-documented. On the other hand, Hoberg \& Philips (2010), Ahern \& Harford (2014), Yan et al. (2016) consider the influence of M\&A events triggered by peer firms and model the inter-dependencies with a different method: Hawkes processes. However, Yan et al. make an unrealistic assumption that: a MA event can effect all the firms in the industry network. This violate what what previous empirial studies suggests, although it solves the sparsity issue.

There are some limitations may restrict the applicability of most existing works. First, although there are several studies that are modeling the inter-dependencies among M\&A events, the representaions of historical events are somehow restrictive due to the limitation of model structures. For example, Hoberg \& Philips (2010) represent the influence from M\&A events triggered by peer firms as a single numerical indicator, while Ahern \& Harford (2014) use several lag terms about "the number of historical M\&A events which previously happened in the local industry network" as the representation of peer effects. Yan et al. (2016) naively assume a historical M\&A event will influence all other firms. They model the inter-dependencies via a fully connected network. Secondly, most existing works are not deal-level prediction tasks. Conceptually, a M\&A event is a mutual agreement between both the acquirer and the target (Blake et al., 1985). The predictions made models which only consider one side of the deal are unreliable due to the lack of consideration of another side of the deal. This issue may also bring error to the estimation of the coefficients for empirical analysis because of the ignorance of related variables: the error term will no longer be a white noise. In the real-world scenarios, the single-side predictive works are hard to be applied. For example, for a bidder company that is seeking potential target companies, a single-side prediction model that predicts “if a firm will be involved into a M\&A deal as target in the future” is not able to give a set of target candidates which are specifically generated for this bidder company. On the other hand, for a startup company that is seeking potential buyers, a single-side prediction model that predicts “if a firm will be involved into a M\&A deal as acquirer in the future” is also not able to help the startup to find potential bidders who are potentially targeting it. Meanwhile, to the best of our knowledge, the only deal-level predictive work (Yang et al., 2014) is a technological M\&A prediction model. The model design and the variable choices are specific for the technology industry, and it is hard to generalize to other industries. Finally, in practice, companies usually require deal-level fine-grained predictions which do not require ad-hoc feature engineering for making predictions in an arbitrary time window. A predictive model that is able to make fine-grained predictions of the rivals' M\&A behavior allows the firm to make prompt decisions and reactions to maintain its competitive advantage in the industry market. However, most existing works require ad-hoc truncations of observation window. Finally, existing dynamic link prediction models are hard to be applied to M\&A since the M\&A event induced network is large-scale, dynamic and sparse. For example, GCN, GraphSAGE are hard to apply to dynamic graphs. NRI is hard to implement on a large-scale network. Dyrep requires dense interactions between nodes to learn meaningful node representations and therefore to make good link predictions.

Given the limitations of existing M\&A predictive models and dynamic link prediction methods, we propose a novel M\&A prediction model based on Temporal Dynamic Industry Network (TDIN). Our method can make fine-grained deal-level predictions, model the rich inter-dependencies among historical M\&A events, and overcome the sparsity issue of M\&A events. Specifically, we parameterize the intensity function of a temporal point process by a deep learning method. Since temporal point processes are a continuous-time model compared with RNNs and logistic regressions, we are able to make fine-grained time prediction. In the meanwhile, temporal point processes have the capability to model the sparsity of the event sequence into its intensity function, which helps us overcome the sparsity issue of M\&A events without making any assumption: we do not need to implement resampling techniques as Yang et al. (2014) do. In fact, a resampling process is almost necessary if acquisition likelihood models are applied toe make deal-level M\&A predictions. We further borrow the idea of graph neural netowk and applied it to make the target predictions specific to a focal acquirer.

\vspace{20pt}

\section{Method}

Our predictive model is composed with two parts: timing module and choice module. Timing part mainly work for predicting the time of next M\&A event triggered by the acquirers. It is determinated by the M\&A intensity function of acquirers. The choice module produces the prediction of a target candidate set that contains the firms with highest probability that they will be acquired by the focal acquirer. Combining the choice part with the timing part, our novel deal-level M\&A method based on TDIN outperforms acquisition likelihood models on a same predictive task. Our method is also able to make fine-grained deal-level predictions that acquisition likelihood models are not able to.

\begin{figure}
    \centering
    \includegraphics[scale=0.3]{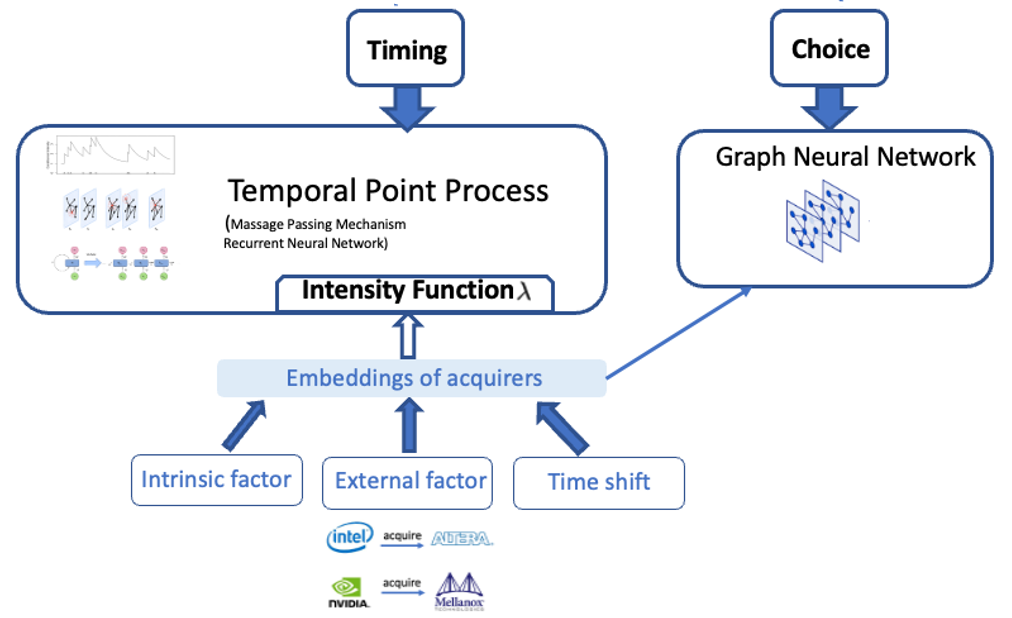}
    \caption{The Overall Structure of Proposed Temporal Dynamic Industry Network M\&A Predictive Model}
    \label{fig:my_label}
\end{figure}

\subsection{Problem Formulation}

We first formally define the deal-level M\&A prediction problem. Let $\mathbb{D} = \{D_d\}$, $d=1,2,3,...,|\mathbb{D}|$, be a set of all frequent acquirer firms, where $|\mathbb{D}|$ is the total number of frequent acquirers in the market. Let $\mathbb{V} = \{\mathbb{V}_v \}$, $v=1,2,3,...,|\mathbb{V}|$, be a set of all target firms over the observation window, where $\mathbb{V}$ is the total number of target firms. However, unlike the acquirer set, the available target firms in the market varies over time. The available target firms in timestamp $t$, $\mathbb{V}^t$, is a subset of $\mathbb{V}$. The relationship among $\mathbb{V}$, $\mathbb{D}$, and $\mathbb{V}^t$ can be shown in Figure 3. $\mathbb{D}$ is a subset of $\mathbb{V}$, $\mathbb{V}^t$ is a subset of $\mathbb{V}$. $\mathbb{D}$ and $\mathbb{V}^t$ may have overlapping with each other.

\begin{figure}
    \centering
    \includegraphics[scale = 0.2]{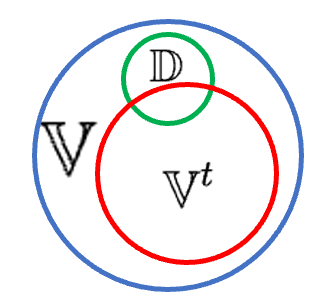}
    \caption{The relationship among $\mathbb{V}$, $\mathbb{D}$, and $\mathbb{V}^t$}
    \label{fig:my_label}
\end{figure}

% For an acquirer $d$, the M\&A events triggered by it is a sequence that each event arrives at timestamps $t_{n^d}$, where $n=1,2,...,N_d$. $N_d$ is the total number of M\&A events triggered by $d$ over the observation window. Assuming no more than one event can happen at an infinitesimal time interval (roughly speaking, at the same time), a timestamp $t_{n^d}$ in the M\&A event sequence of $d$ uniquely corresponds to an available target firm (that $d$ acquired at $t_n^d$) in $\mathbb{V}^{t_n^d}$. We denote this target firm in $\mathbb{V}$ ($\mathbb{V}^{t_n^d} \subseteq \mathbb{V}$) as $\mathbb{V}_{v_{t_n^d}}$, where $v_{t_n^d} \in \{1,2,3,...,|\mathbb{V}|\}$. Back in the previous example, $\mathbb{D}_d$ is AMD (AMD is the $d_{th}$ element in the set $\mathbb{D}$), $t_{n^d}$ is April 7, 2021, which represent this M\&A event is the $n_{th}$ event in the M\&A event sequence of AMD, and it happened at April 7, 2021. $\mathbb{V}_v_{t_n^d}$ is XILINX, which is saying XILINX is one of firms in $\mathbb{V}^{t_n^d}$ and also the $v_{t_n^d}\;th$ element in $\mathbb{V}$.

For an acquirer $d$, the M\&A events triggered by it is a sequence where each event arrives at timestamps $t_n^d$, where $n=1,2,...,N_d$. $N_d$ is the total number of M\&A events triggered by $d$ over the observation window. Assuming no more than one event can happen at an infinitesimal time interval (roughly speaking, at the same time), a timestamp $t_n^d$ in the M\&A event sequence of $d$ uniquely corresponds to an available target firm (that $d$ acquired at $t_n^d$) in $\mathbb{V}^{t_n^d}$. We denote this target firm in $\mathbb{V}$ ($\mathbb{V}^{t_n^d} \subseteq \mathbb{V}$) as $\mathbb{V}_{v_{t_n^d}}$, where $v_{t_n^d} \in \{1,2,3,...,|\mathbb{V}|\}$. Back in the previous example, $\mathbb{D}_d$ is AMD (AMD is the $d_{\text{th}}$ element in the set $\mathbb{D}$), $t_n^d$ is April 7, 2021, which represents this M\&A event is the $n_{\text{th}}$ event in the M\&A event sequence of AMD, and it happened on April 7, 2021. $\mathbb{V}_{v_{t_n^d}}$ is XILINX, which is saying XILINX is one of the firms in $\mathbb{V}^{t_n^d}$ and also the $v_{t_n^d}\;\text{th}$ element in $\mathbb{V}$.

Given the history of M\&A events until time $t$, $\mathfrak{H}_t$. For an acquirer $\mathbb{D}_d$, a deal-level M\&A prediction problem has two predictive objectives: the time of the acquirer's next merger and acquisition event, and the target of this event.

\vspace{10pt}

\subsection{Temporal Dynamic Industry Network (TDIN)}

The key idea of our method is to evaluate the “incentives” of acquirers to trigger a M\&A event with each available target in the public market. Borrowing the idea of the intensity function from the point process, such incentives could be quantified as the value of an intensity function $\lambda$. For an acquirer $d$, to answer its incentives to acquire target $v$ at time $t$, we need to compute the value of $\lambda_d(t, v)$. Our method focuses on this value and further decomposes $\lambda_d(t, v)$ into two parts to capture the timing and choice respectively: we name them the timing module and the choice module.

$$\lambda_d(t, v| \mathfrak{H}_t) = \lambda_d(t|\mathfrak{H}_t) \times P_{d}(v|t)$$

where $\lambda_d(t|\mathfrak{H}_t)$ capture the overall incentive to acquire for the focal acquirer, which varies over time. This part is modeled by the timing module. The overall incentive is further distributed to each possible target at time $t$ as $\sum_v P_{d}\left(v|t  \right) = 1, \;v\in \mathbb{V}^t$.

\subsubsection{Timing Module}

First, we define the M\&A conditional intensity function (CIF) that describes the incentive of mergers to trigger a merger and acquisition event:

\begin{equation*}
\lambda_d(t)=\lim _{\Delta t \rightarrow 0} \frac{\mathbb{E}\left(N(t+\Delta t)-N(t) \mid \mathfrak{H}_{t}\right)}{\Delta t}  
\end{equation*}

The incentive is measured by the average number of new M\&A events triggered by an acquirer $\mathbb{E}\left(N(t+\Delta t)-N(t) \mid \mathfrak{H}_{t}\right)$ in an infinitesimal time slot $\Delta t$, which analogizes the definition of intensity function of a temporal point process. To model the complex dependencies between the intrinsic factors, external factors and the incentive to merge, we parameterize it by the following way:

$$\lambda_{d}(t)=f_{\lambda}\left(w_{e}^{\top} e_{d}^{t}+w_{c}^{\top} c_{d}^{t}+\omega\left(t-t_{d}^{-}\right)\right)$$

where $e_d^t \in \mathbb{R}^{d_1}$ is a vector representation of the intrinsic factors. This term considers the inner initiative from the focal acquirer such as liquidity adequacy (Ragothaman et al., 2003; Ali-Yrkko et al., 2005), and other available financial and managerial variables suggested by previous empirical research. $c_d^t \in \mathbb{R}^{d_2}$ is a vector representation of the extrinsic factors, which cumulatively summarize the peer effect from the past M\&A events. Vector $w_e \in \mathbb{R}^{d_1}$ contains learnable parameters that share the same size as $e_d^t$, while $w_c \in \mathbb{R}^{d_2}$ is also a learnable parameter. The summation $w_{e}^{\top} e_{d}^{t}+w_{c}^{\top} c_{d}^{t}$ is inspired by “the base rate” of Hawkes process (Hawkes, 1971), and models the spontaneous rate of the focal acquirer. $\omega$ is a simple monotonic decreasing function to model time shifts (Du et al., 2016) between two events in the timeline:

$$\omega=w_{o_1} * e^{-w_{o_2}(\Delta t)}$$

where $w_{o_1}$ and $w_{o_2}$ are two learnable scalars.

For an acquirer $d$, we parameterize its conditional intensity function over time. So, statistically speaking, we are modeling M\&A events as a multi-dimensional point process, while each dimension is a timeline of an acquirer with different types of events: self-triggered M\&A events, peer effect, and updates of its intrinsic factors. Mathematically speaking, conditional intensity function for each firm is a piecewise continuous function, where the jumps on the non-differentiable points are parameterized by a deep learning algorithm, and the derivable intervals between those points are parameterized by a differentiable monotonic decreasing function with time.

\begin{figure}
    \centering
    \includegraphics[scale = 0.35]{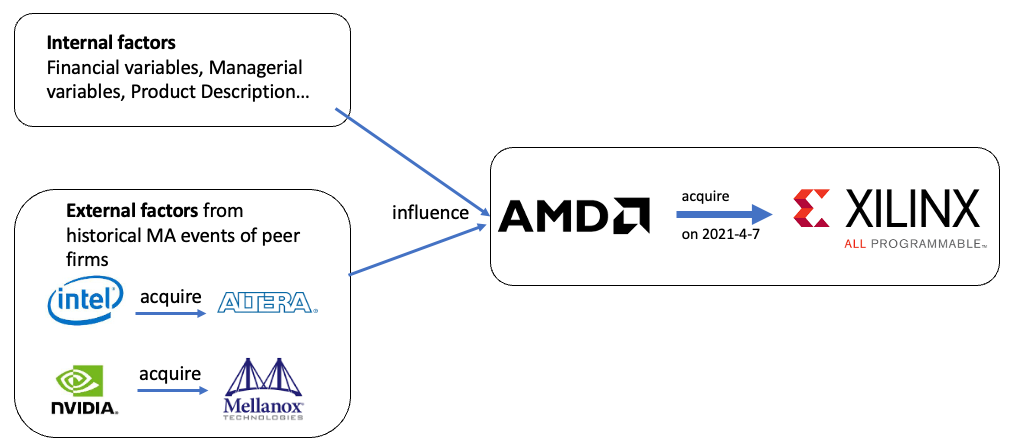}
    \caption{Intrinsic and Extrinsic Factors of Merger and Acquisition Events: An Example}
    \label{fig:my_label}
\end{figure}

Specifically, for the vector representation of intrinsic factors $e_d^t$, we consider two parts of the information as the input: accounting variables and textual embeddings from 10-K\footnote{only the “product description” part of 10-K}. The reasons why we consider textual information are: first, recent studies have shown that textual information from 10-K has significant predictive power for M\&A events (Hoberg \& Philips, 2010; Hoberg \& Philips, 2016; Routledge et al., 2017). Secondly, one of the most important motivations of firms continuing to conduct M\&A transactions is that they are seeking product synergy to maintain competitive advantage (Hoberg \& Philips, 2010). However, accounting variables have no capability to provide adequate information about firms’ products. In the meanwhile, to the best of our knowledge, there is no natural language processing technology that can process table information in 10-K documents, which means most numerical values of financial variables are completely ignored in textual analysis. So, textual information and financial variables complement each other very well.

We first create textual embeddings of the focal acquirer’s “Business Description” section of their 10-K filings $text_d^t \in \mathbb{R}^{\frac{d_1}{2}}$. We use a pre-trained natural language processing model (NLP): FinBERT (Araci \& Dogu, 2019) to generate textual embeddings.The available financial variables are first input into a fully-connected neural network (FCN), to generate the financial embedding $fin_d^t \in \mathbb{R}^{\frac{d_1}{2}}$. Those two embeddings are then concatenated together as  $e_d^t$.

\begin{figure}
    \centering
    \includegraphics[scale=0.5]{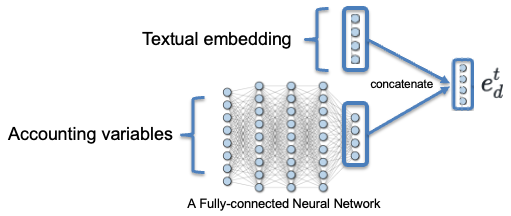}
    \caption{The $e_d^t$ Embedding Network}
    \label{fig:my_label}
\end{figure}

For the vector representation of extrinsic factors $c_d^t$, we introduced a dynamic industry network with pre-defined network structures. Unlike most of the dynamic networks, the network structures of our dynamic network is not determined by interactions between nodes but pre-defined by TNIC similarity\footnote{TNIC: Text-based Network Industry Classifications.}. We denote the dynamic industry network as $G_t =(\mathcal{V}_t, \mathcal{E}_t)$, where the set of nodes $\mathcal{V}_t$ contains all public firms in the market at time $t$. For edge set $\mathcal{E}_t$, we only keep the edges connecting with top ten peer firms for each focal firm, while a higher TNIC similarity represents higher closeness between firms. Since TNIC similarity is also varying over time, the edge set also have subscript $t$. The pre-defined edges in the dynamic industry network allow us to retain message passing mechanisms in the graph neural network while alleviating the sparsity issue of M\&A events. The message passing mechanisms mimic the propagation of the peer effect among industry networks (Ahern \& Harford, 2016). These help our model to capture the rich inter-dependencies among M\&A events.

\begin{figure}
    \centering
    \includegraphics[scale=0.3]{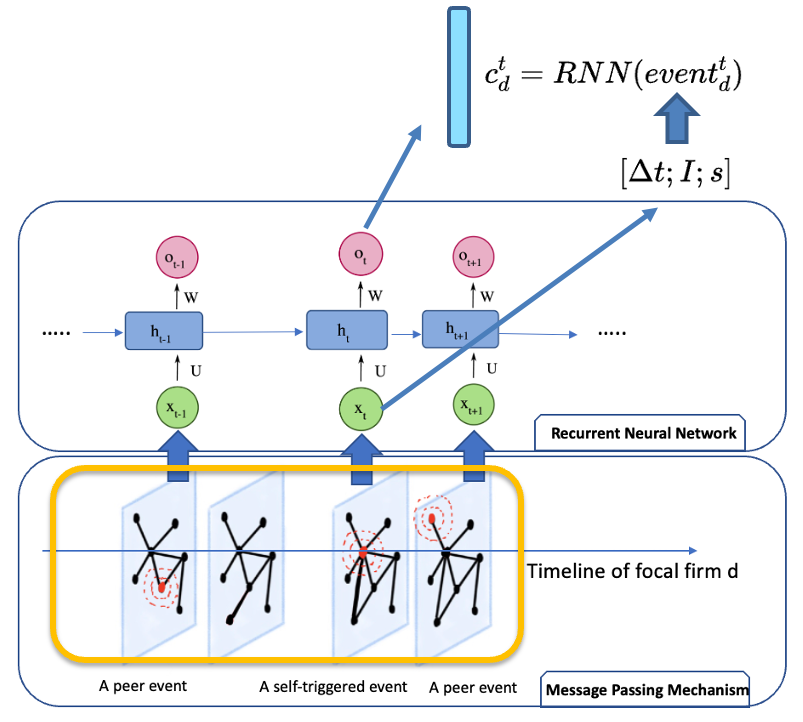}
    \caption{The $c_d^t$ Embedding Network}
    \label{fig:my_label}
\end{figure}

On the top of the dynamic industry network, we further introduced recurrent neural networks to model the inter-dependencies among M\&A events. Specifically, for an acquirer, an event embedding $event_d^{t} \in \mathbb{R}^{3}$ is created as the vector representation of the information of the event happening in the timestamp $t$ in the focal acquirer $d$’s timeline. The four elements of an event embedding are: time difference $\Delta t$ between last event and the focal event, event type $I$, TNIC (Hoberg \& Philips, 2016) \footnote{see https://hobergphillips.tuck.dartmouth.edu/industryclass.htm for detail} similarity $s$ between the acquirer and the target. There are two types of event that we consider in the event embeddings: self-triggered M\&A event (represented by an integer 0) and peer effect (represented by an integer 1). For example, on the timeline of focal acquirer $d$, at timestamp $t$, a peer firm triggers a M\&A event as acquirer. First, we look up what is the timestamp that the most recent event happened before this event, and compute the time difference $\Delta t$ between them. Then, since this is a M\&A event triggered by a peer firm, the event type will be embedded as an integer value “1”.  Finally, the TNIC similarity value is looked up in the TNIC dataset, and a corresponding event embedding is created. The event embedding will then be passed into a recurrent neural network as input. 

\begin{figure}
    \centering
    \includegraphics[scale=0.3]{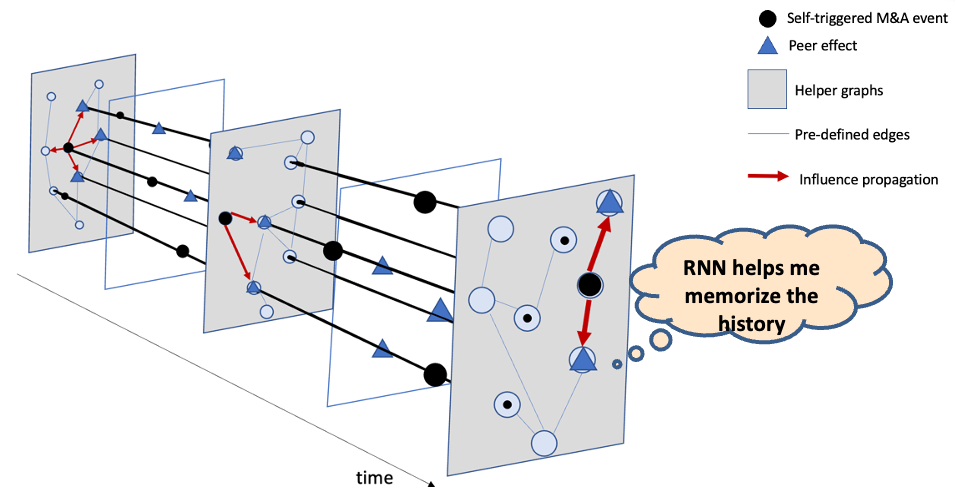}
    \caption{The Overall Structure of The Timing Module}
    \label{fig:my_label}
\end{figure}

Given the big picture of the timing module as figure 7, if we view it in order of time, from past to recent, the multiple lines are a multi-dimensional point process while each line models the M\&A behaviors of each acquirer. If we view it cross-sectionally, each slice is an industry network that directs the path of the propagation of peer effects. Different slices show different graph structures since the industry network is a dynamic network: the relationship between firms varies over time, which changes the direction of the propagation. In addition, for each acquirer, the RNN\footnote{the RNN shares parameters over all acquirers} helps remember the peer effect from the past.

\vspace{10pt}

\subsubsection{Choice Module}

In the choice module, we borrow the idea of graph neural network, use the message passing and aggregation function to mimic the process that firms are actively collecting information about their rivals, and then make prompt reaction which may leads change of their own investment decisions (Bernard et al., 2020; Bustamante \& Fresard, 2021). Like what we do in the timing module, the network structures are pre-defined by TNIC similarity. For the initial node embedding in the industry network, we reused the fully connected network that we introduce when creating the vector representation of the intrinsic factors $e_d^t$. This makes the timing and choice module more cohesive with each other, and allow more gradient pass during the backpropagation process, which is the key of parameter learning for deep learning methods (Rumelhart et al., 1986a). Also, the recent studies (Bernard et al., 2020) indicate that the information we embedded into $e_d^t$ is part of the information firms are monitoring about their rivals.

\begin{figure}
    \centering
    \includegraphics[scale=0.4]{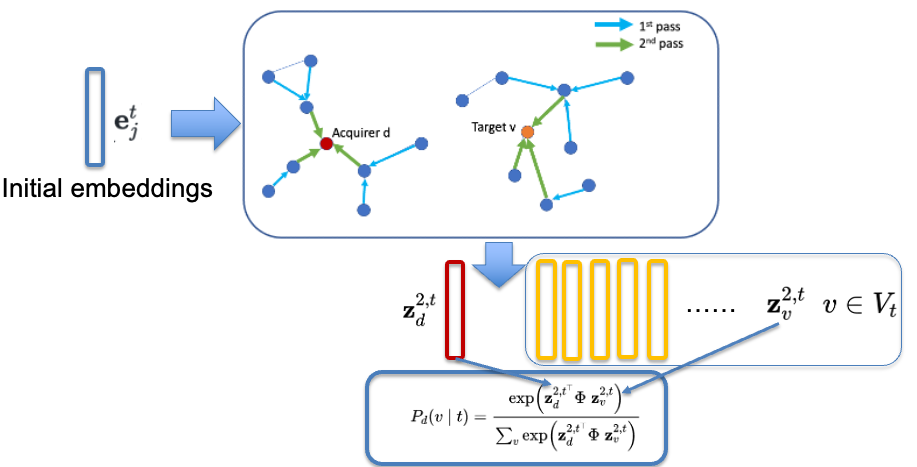}
    \caption{The Choice Module}
    \label{fig:my_label}
\end{figure}

After putting the initial embeddings into the choice module, the information will pass through two layers of message passing procedures and aggregations, which allows the acquirers and targets to grasp their peer firms information:

\begin{center}
\begin{align*}
    & \text{1st pass} \left\{
    \begin{aligned}
        \forall j: \mathbf{a}_{j}^{1, t} &= f_{\text{agg}}^{1}\left(\left\{e_u^t \;\;\forall u \in \mathcal{N}(j)   \right\}\right) \\
        \forall j: \mathbf{h}_{j}^{1, t} &= f_{\text{comb}}^{1}\left(e_u^t, a_j^{1,t}\right)
    \end{aligned}
    \right. \\
    & \text{2nd pass} \left\{
    \begin{aligned}
        \forall j: \mathbf{a}_{j}^{2, t} &= f_{\text{agg}}^{2}\left(\left\{h_u^{1,t} \;\;\forall u \in \mathcal{N}(j)   \right\}\right) \\
        \forall j: \mathbf{z}_{j}^{2, t} &= f_{\text{comb}}^{2}\left(\mathbf{h}_j^{1,t}, a_j^{2,t}\right) \\
    \end{aligned}
    \right.
\end{align*}
\end{center}

\vspace{20pt}

The output of the choice module is the new node embeddings of each firm in the industry network. Specifically, for the focal acquirer, the output acquirer embedding will be $z^{2,t}_d$. Along with $z^{2,t}_d$, there will be a set of target embeddings $z_v^{2,t}$, where $v \in \mathbb{V}^t$. We further take $z_d^{2, t}$ and each $z_v^{2,t}$ into a softmax layer to compare the compatibility between the focal acquirer $d$ and each $v \in \mathbb{V}^t$:

$$y_v^{(d, t)} = P_{d}\left(v | t \right)=\frac{\exp \left(\mathbf{z}_{d}^{2, t^{\top}} \Phi \mathbf{z}_{v}^{2, t}\right)}{\sum_{v} \exp \left(\mathbf{z}_{d}^{2, t^{\top}} \Phi \mathbf{z}_{v}^{2, t}\right)}$$

The advantage of using a softmax layer is the summation over the compatibility between the focal acquirer $d$ and all other possible target candidates is equal to 1: $\sum_v P_{d}\left(v |t \right) = 1, \;v\in \mathbb{V}^t$.

\subsection{Empirical Evaluation}

\subsection{Data}

Our data is composed of four components. The M\&A events data came from Thomson Reuters' SDC Platinum database\footnote{The complete variable descriptions could be found at http://mergers.thomsonib.com/td/DealSearch/help/def.htm}. Our accounting variables are collected from the Compustat-Capital IQ database. The textual data is the item 1, or “Business Description” section of 10-K form\footnote{See the definition of item 1 at  https://www.sec.gov/fast-answers/answersreada10khtm.html}. We collect them from the SEC\footnote{SEC is short for “U.S. SECURITIES AND EXCHANGE COMMISSION”}. The industry network is built on the Text-based Network Industry Classifications (TNIC) database\footnote{See the complete data description at http://hobergphillips.tuck.dartmouth.edu/}, from Hoberg-Phillips Data Library. Our observation window is from the start of 1997 to the end of 2020.

For the M\&A events data, we first remove all the M\&A deals that happened within a company. They are asset reassignment and rearrangement inside a single firm which we do not consider. We further solely focus on complete and majority takeovers following the standard definition of a “merger and acquisition” (Betton et al., 2008; Ahern et al., 2014; Routledge et al., 2017). Specifically, the definition of “majority” takeover is the following: the percentage of share that acquirer acquire has to exceed 20\% of the total share of the target; the fraction of shares held by the bidder after a completed transaction must be larger than 50\%; the value of the transaction has to surpass a million US dollar; The transaction type could not be “block purchases”, “creeping acquisition”, and “privatization”. Similar to recommendation systems, we are facing cold-start problem (Schein et al., 2002; Lika et al., 2014) for those acquirers who never trigger any M\&A events or only trigger a few. An analogy in recommendation systems in social networks is: if a user never links to any other user, it is nearly impossible to make precise recommendations to this user. Similarly, if a firm has never acquired any firm before, it is very hard to predict which firm will be targeted by it. To alleviate this issue, we only consider the firms who already triggered 4 or more merger and acquisition events during the observation window. We call them “frequent acquirers”, and we have 385 of them. For accounting variables, we majorly follow the choices of two recent M\&A predictive works (Yang et al., 2014; Bernard et al., 2020), but also consider the data availability. Our final choices comprehensively cover the basic operation information for companies. They are variables about the assets of a firm, leverage, current condition of business operation, and the research and development spend\footnote{see Appendix B for detail}. Unfortunately, the missing ratios of the selected variables are still not low. For firms who occasionally have several missing values, we interpolate that value from the previous year. If a variable is totally missing over the entire observation period for a firm, we interpolate the average number of all other firms. For TNIC data that we used to construct the pre-defined edges in the industry network, we obtain the TNIC data from 1996 to 2019 (one year before the observation window). Since TNIC is conceptually a fully connected network with different weights on each edge, we only keep the top ten peers which follows the previous works (Hoberg \& Philips, 2010; Hoberg \& Philips, 2016). However, some industries are highly competitive, and others may not. To allow our network structures to have adaptability for different industries, we also set an absolute threshold for the TNIC similarity. In other words, if TNIC similarity exceeds a certain threshold, we will keep the link in the network while ignoring the former rule.

\subsection{Parameter Estimation}

Given a $|\mathbb{D}|$-dimensional MA sequence of $N_D$ frequent acquirers we observed until $t$, for all frequent acquirers in $\mathbb{D}$, the loss function can be separated into two terms: timing loss and choice loss(Du et al., 2016).

\begin{equation}
    L(S_t) = \sum_{d}\sum_{n} (L_{time}^{(d,n)} + L_{choice}^{(d,n)})
\end{equation}

For the timing loss part, we use the negative log likelihood function:

\begin{equation}
    \sum_{d}\sum_{n} (L_{time}^{(d,n)}) =  \sum_{d} \left[ \sum_{n=1}^{N_{d}} \lambda_d(t_n^d) -\int_{0}^{t_{N_{d}}^d} \lambda_d(\tau) d\tau \right]
\end{equation}

For the choice loss part, we take the choice prediction $\hat{v_{t_n^d}}$ as a binary-class classification problem. So we use binary cross entropy loss:

\begin{equation}
    \sum_{d}\sum_{n}(L_{choice}^{(d,n)})= - \sum_d\sum_{n} \frac{1}{|V^{t_n^d}|} \sum_{v \in V^{t_n^d}}  [y_v^{t_n^d} log(\hat{y}_{v}^{t_n^d}) + (1-y_v^{t_n^d}) log(1-\hat{y}_v^{t_n^d})]
\end{equation}

\subsection{Evaluation}

We benchmark the performance of our proposed method to the most widely-used model in the existing M\&A predictive works: Acquisition Likelihood Model (Palepu, 1986).

Besides all the accounting variables we used, our proposed model also takes top ten TNIC similarity and past M\&A events into account. We add the average number of TNIC similarity of top ten peers as a predictor in the acquisition likelihood model. Also, following the work of Ahern \& Harford (2014), we add three lag terms to represent the historical merger and acquisition events that happened in the local environment of the focal acquirers. 

Acquisition likelihood model could only make predictions for a fixed time period, which requires ad-hoc time truncation. For example, most previous works truncate time year by year, and make predictions one year after the observation period. However, our model is built based on a temporal point process which could make continuous-time prediction. 

Given the historical M\&A events until time $t$, $\mathfrak{H}_t$, the latest available accounting variables at time $t$, and the latest network structures determinated by TNIC data at $t$, we could compute the value of the M\&A intensity functions at timestamp $t$ for a focal acquirer $d$. Since the relationship between the intensity function and the probability density function $p_d(t)$ is fixed as equation (1), we first make a precise time prediction as the expectation of the distribution of intensity at timestamp $t$  for acquirer $d$: $\lambda_{d}(t)$:

\begin{center}
 \begin{equation*}
    \widehat{t}_{d}^{+}=\mathbb{E}\left[t \mid \mathfrak{H}_{t}\right]=\int_{t}^{\infty} t\; p_{d}(t) d t
\end{equation*}   
\end{center}

Given the next M\&A event time triggered by acquirer $d$, we then convert the continuous time prediction to discrete one which matches the predictive label of existing acquirer-side prediction. Specifically, if $\widehat{t}_{d}^{+}$ fall inside the next year window, then we marked the predictive label as “1”, otherwise marked as “0”.

Following the procedure above, we use the merger and acquisition data from 1997 to 2019 as our training set the acquisition likelihood model achieves AUC score 0.5451, while our model achieve AUC score 0.581, which yields a 6.6\% increase of performance. 

\begin{table}[h]
\centering
\caption{Performance Comparison of M\&A Prediction Models}
\label{tab:performance_comparison}
\begin{tabular}{lcc}
\hline
\textbf{Model} & \textbf{AUC Score} & \textbf{Performance Increase} \\
\hline
Acquisition Likelihood Model (Palepu, 1986) & 0.5451 & - \\
Proposed TDIN Model & 0.5819 & 6.6\% \\
\hline
\end{tabular}
\end{table}

\section{Discussion and Analysis}

To comprehensively evaluate the effectiveness of our proposed Temporal Dynamic Industry Network (TDIN) model for M\&A prediction, we conducted an ablation study by systematically removing or modifying key components of the model. This section combines the ablation study results with an in-depth discussion, providing insights into the contributions of each component and their implications for M\&A prediction.

\subsection{Ablation Study Results}

Table~\ref{tab:ablation_study} summarizes the performance impact of each component in our model. We report the AUC scores and the percentage change in performance when each component is ablated.

\begin{table}[h]
\centering
\caption{Ablation Study Results}
\label{tab:ablation_study}
\begin{tabular}{lcc}
\hline
\textbf{Model Variant} & \textbf{AUC Score} & \textbf{Performance Change} \\
\hline
Full TDIN Model & 0.5819 & - \\
\hline
\textit{Component Ablations:} & & \\
\quad Without Textual Embeddings & 0.5603 & -3.71\% \\
\quad Without Dynamic Industry Network & 0.5492 & -5.62\% \\
\quad Without Peer Effect Modeling & 0.5557 & -4.50\% \\
\quad Without RNN in Timing Module & 0.5685 & -2.30\% \\
\quad Without GNN in Choice Module & 0.5541 & -4.78\% \\
\hline
\end{tabular}
\end{table}

\subsection{Analysis of Model Components}

\subsubsection{Impact of Textual Embeddings}

Removing the textual embeddings derived from 10-K filings resulted in a 3.71\% decrease in AUC score. This confirms the importance of incorporating unstructured textual information to capture qualitative factors such as product synergies, strategic intentions, and managerial outlooks that are not fully reflected in financial variables. Textual analysis allows the model to understand nuanced information, which is critical in M\&A decisions where qualitative assessments often play a significant role.

\subsubsection{Importance of Dynamic Industry Network}

The dynamic industry network, modeling the evolving relationships between firms, proved to be the most critical component, with its removal causing a significant 5.62\% drop in performance. This highlights the necessity of capturing the temporal dynamics of inter-firm relationships and the competitive landscape. By utilizing a dynamic network, the model can account for shifts in industry structure, entry and exit of firms, and changing competitive pressures, which are vital factors influencing M\&A activity.

\subsubsection{Contribution of Peer Effect Modeling}

Excluding the peer effect component led to a 4.50\% decrease in performance, emphasizing the influence of recent M\&A events by peer firms on a focal firm's M\&A decisions. This finding aligns with empirical studies indicating that firms are more likely to engage in M\&A activities following similar actions by competitors due to factors like competitive rivalry, industry trends, and herding behavior. Modeling peer effects allows the TDIN to capture these contagion effects, improving predictive accuracy.

\subsubsection{Role of RNN in Timing Module}

The recurrent neural network (RNN) in the timing module contributed to a 2.30\% improvement in performance. While its impact is relatively modest compared to other components, the RNN effectively captures temporal dependencies and patterns in the timing of M\&A events. This temporal modeling is essential for understanding not just whether a firm will engage in M\&A, but also when it is likely to occur, which is crucial for dynamic prediction tasks.

\subsubsection{Significance of GNN in Choice Module}

The graph neural network (GNN) in the choice module is a key component, with its removal resulting in a 4.78\% decrease in performance. The GNN enables the model to perform message passing and aggregation over the dynamic industry network, allowing firms to collect and react to information about their rivals. This mechanism effectively captures the interdependencies and strategic interactions among firms, which are fundamental in M\&A decision-making processes.

\subsection{Overall Performance Improvement}

The proposed TDIN model achieves a 6.6\% improvement in AUC score compared to the traditional Acquisition Likelihood Model, demonstrating the effectiveness of integrating dynamic network structures, textual information, and advanced deep learning techniques in M\&A prediction. The significant performance gain underscores the model's ability to capture the complex, multifaceted nature of M\&A activities.

\subsection{Implications for M\&A Prediction and Research}

The findings from our ablation study have several implications:

\begin{itemize}
    \item \textbf{Holistic Modeling Approach:} The substantial performance drops observed when key components are removed indicate that M\&A prediction benefits from a holistic modeling approach that integrates financial data, textual information, network dynamics, and temporal patterns.
    \item \textbf{Importance of Dynamic Networks:} Modeling the industry as a dynamic network is crucial for capturing the evolving competitive landscape and strategic interactions among firms. This approach can be extended to other domains where inter-entity relationships play a significant role.
    \item \textbf{Advancement in Deep Learning Applications:} The effective use of RNNs and GNNs highlights the potential of deep learning architectures in handling complex temporal and relational data in financial prediction tasks.
\end{itemize}

\subsection{Limitations and Future Research}

While the TDIN model shows promising results, there are limitations to consider:

\begin{itemize}
    \item \textbf{Data Availability:} The model relies on comprehensive financial and textual data, which may not be available for all firms, particularly smaller or private entities.
    \item \textbf{Model Complexity:} The integration of multiple deep learning components increases computational complexity and may pose challenges for scalability and real-time prediction.
    \item \textbf{Interpretability:} Deep learning models often lack transparency, which can be a concern in financial contexts where explainability is important for decision-makers.
\end{itemize}

Future research could address these limitations by exploring methods for simplifying the model without significant loss of accuracy, enhancing interpretability through techniques like attention mechanisms, and extending the model to incorporate additional data sources such as news articles or social media sentiment.

\section{Conclusion}

In this paper, we develop a novel M\&A predictive model based on a deep learning method. Our proposed method has several novelties and advantages. First, it can effectively capture the rich inter-dependencies among historical M\&A events. Specifically, addressing the research gap that most existing works fail to model the complex inter-dependencies because of their restrictive representations. Second, our model is capable of making deal-level predictions without restriction of the industry domains. Third, our model also can make continuous time prediction which does not require ad-hoc feature engineering to transform continuous time M\&A events to fixed-length features, and avoid the loss of precise information during such transformations. We demonstrate the superiority of our proposed method over the acquisition likelihood model from most existing methods.

\pagebreak

\nocite{*}
\printbibliography

\pagebreak 
\appendix

\section*{Appendix}

\subsection*{Appendix A}

\textbf{Proof of equation (1)}:

Assume we have observed $n$ times of observations until current time point $t_{c}$ (not including) of M\&A events, which happened in $t_1, t_2, t_3,..., t_n$. Assume the value of probability density function $f$ (likelihood) in those timestamps are $f(t_1), f(t_2), f(t_3), ..., f(t_n)$. Then, we have the cumulative likelihood for all the observed events in the form:

$$L = \prod_{i=1}^{n} f(t_i)$$

Recall the definition of intensity function: the instantaneous rate of occurrence of events:

$$\lambda(t)=\lim _{d t \rightarrow 0} \frac{\mathbb{E}\left(N(t+d t)-N(t) \mid \mathfrak{H}_{t}\right)}{d t} = \lim _{d t \rightarrow 0} \frac{\operatorname{Pr}(N(t+d t)-N(t)=1)}{d t}$$

Here $\lambda$ could be seen as the 1st derivitive of a probability density function (Keep in mind that $\lambda$ itself is neither a probability nor a value of proability density function).

Assume $T$ is the timestamp that the next event arrives. For the numerator, it is the joint probability that $T$ is in the interval $[t, t+dt)$ and $T\le t$:

$$P(t \leq T<t+d t, ; T \geq t)=P(t \leq T<t+d t)$$

when $dt$ is an infinitesimal interval:

$$P(t \leq T<t+d t)=\int_{t}^{t+d t} f(x) d x \approx f(t) \times d t$$

For the denumerator:

$$P(T \geq t)=S(t)=1-F(t)$$

Overall, the intensity function could be writen as:

$$\lambda(t)=\frac{\frac{f(t) d t}{1-F(t)}}{d t}=\frac{f(t)}{1-F(t)}=\frac{f(t)}{S(t)}$$

Assume the cumulative density function is $F$,  $f(t) = \frac{\mathrm{d} F(t)}{\mathrm{d} t}$. Then, the intensity function can be solely represented by $F$:

$$\lambda(t) = \frac{f(t)}{1-F(t)} = \frac{\frac{\mathrm{d} F(t)}{\mathrm{d} t}}{1-F(t)} = -\frac{d log(1-F(t))}{dt}$$

We further take the integral over the timestamp of the most recent event $t_n$ and current timestamp $t$:

\begin{equation}
  \int_{t_n}^{t} \lambda(\tau) d\tau = log(1-F(t_n)) - log(1-F(t))  
\end{equation}

We have already pass the timestamp $t_n$, while the density function $f$ or $F$ is w.r.t. the time that next event (the $n+1_{th}$ event) will happen. This means the value of density function of the time of $n+1_th$ event $T$ will be zero everywhere before $t_n$. So we could simplify equation (3) to:
\begin{equation}
  \int_{t_n}^{t} \lambda(\tau) d\tau = - log(1-F(t))  
\end{equation}

rearrange equation (6), we get the relationship between the probability density function $f$ and the intensity function $\lambda$:

$$f(t)= \lambda(t) \exp(-\int^{t}_{t_n}\lambda(\tau) d\tau)$$

\subsection*{Appendix B}

The choice of accounting variables mainly follows two recent M\&A predictive studies (Yang et al., 2014; Bernard et al., 2020). Yang et al. (2014) mentioned many financial, managerial and other variables\footnote{see table 2 of (Yang et al., 2014) for more detail}. For financial variables, they are: size, market-to-book ratio, cash, return on assets(ROA), sales to total assets (assets turnover), debt-to-equity ratio (leverage), current ratio, return on equity(ROE), debt-to-assets, capital expenditures to total asset, capital expenditures to total asset growth, sale growth, cash flow to total asset ratio, cash flow to sale ratio, asset growth. For managerial variables, they are: management inefficiency, resource richness, industry variations, relevance degree of business boundaries, export orientation, age of the firm. For other variables, they are: cost-to-income ratio, cost of goods sold divided by inventory, outperformance, tax shield effects, ratio of tangible assets to total assets, return on investment, profit margin, average dividend for last three years, shareholders, earnings before interest and taxes, operating income after depreciation, common shares traded divided by common shares outstanding. Bernard et al. (2020) mentioned seven accounting variables, the only one that was not mentioned by Yang et al. (2014) is PPE\footnote{see definition at Appendix A of (Bernard et al., 2020) }. However, due to the restriction of data availability, we only adopt part of the variables mentioned above. The variable we adopted, the formula for creating those variables, and the brief definitions of those variables are listed below\footnote{The abbreviations follow Compustat.}.

\begin{center}
\begin{landscape}
\begin{table}[!ht]
    \centering
    \begin{tabular}{|l|l|l|}
    \hline
        \textbf{Variable} & \textbf{Formula} & \textbf{Definition} \\ \hline
        ~ & \textbf{Bernard et al. (2020)} & ~ \\ \hline
        Size & $at$ & A firm’s total assets \\ \hline
        Market-to-book ratio & $(at+prcc\_f*csho-ceq-txdb)/at$ & Market-to-book assets ratio \\ \hline
        Leverage & $(dlc+dltt)/at$ & Book leverage \\ \hline
        ROA & $ib/at$ & Return-on-assets \\ \hline
        Sales growth & $(sale\_\{t\}-sale\_\{t-1\})/sale\_\{t-1\}$ & Sales growth \\ \hline
        PPE & $ppent/at$ & A firm’s net plant, property, and equipment, scaled by total assets \\ \hline
        Cash & $ch$ & Cash \\ \hline
        ~ & \textbf{Yang et al (2014)} & ~ \\ \hline
        Sale & $sale$ & A firm’s net sales \\ \hline
        Cash-to-asset ratio & $ch/at$ & Cash flow to total assets ratio \\ \hline
        Cash-to-sales ratio & $ch/sale$ & Cash flow to sales ratio \\ \hline
        Sales-to-asset ratio & $sale/at$ & Net sales/total assets \\ \hline
        Current ratio & $act/lct$ & Current assets scaled by its current liabilities \\ \hline
        Asset growth & $(at\_\{t\}-at\_\{t-1\})/at\_\{t-1\}$ & Total asset growth \\ \hline
        GSI & $cogs/invt$ & Cost of goods sold divided by inventory \\ \hline
        DE & $(dlc+dltt)/CEQ$ & Debt to common equity \\ \hline
        R$\backslash$\&D & $rdip$ & In process R$\backslash$\&D expense \\ \hline
        ROE & $ib/ceq$ & Return on equity \\ \hline
    \end{tabular}
\end{table}

\end{landscape}
\end{center}

\subsection*{Appendix C}

Figure 9 displays the visualization of the adjacency matrix in 1997 after pre-processing the TNIC data. Each colored point in the picture is a link. On the contrary, a blank point at $(i, j)$ represents that there is no link between firms $i$, $j$,

\begin{figure}
    \centering
    \includegraphics[scale=0.3]{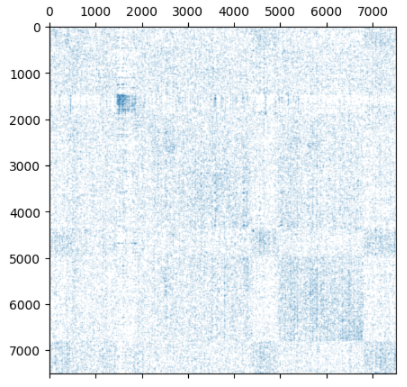}
    \caption{Visualization of the Adjacency Matrix in 1997}
    \label{fig:my_label}
\end{figure}

\end{document}